\documentclass[preprint]{aastex}
\usepackage{emulateapj5}

\input epsf.sty
\usepackage{graphicx}
\usepackage{times}
\def\l{$\lambda$}

\def\rfe{R$_{\rm FeII}$}

\def\ltsima{$\; \buildrel < \over \sim \;$}
\def\simlt{\lower.5ex\hbox{\ltsima}}            
\def\gtsima{$\; \buildrel > \over \sim \;$}

\def\simgt{\lower.5ex\hbox{\gtsima}}            

\def\civ{{\sc{Civ}}$\lambda$1549\/}

\def\civbc{{\sc{Civ}}$\lambda$1549$_{\rm BC}$\/}

\def\cm3{cm$^{-3}$\/}
\def\hb{{\sc{H}}$\beta$\/}

\def\hbbc{{\sc{H}}$\beta_{\rm BC}$\/}
\def\hbnc{{\sc{H}}$\beta_{\rm NC}$\/}

\def\oiiiopt{{\sc{[Oiii]}}$\lambda\lambda$4959,5007\/}
\def\o4363{{\sc{[Oiii]}}$\lambda$4363\/}

\def\feii{{Fe\sc{ii}}$_{\rm opt}$\/}
\def\fe{{\sc{Fe}}\/}
\def\dvr{{$\Delta$v$_r$}}
\def\vr{{v$_r$}}

\def\fe76087{{\sc [Fe vii]}$\lambda$6087\/}
\def\oiii{{\sc [Oiii]}$\lambda$5007}

\def\kms{km~s$^{-1}$}

\usepackage{graphicx}
\input epsf.sty
\usepackage{graphics}
\usepackage{times}
\submitted{ accepted July 11, 2002}

\begin{document}
\title{Kinematic Linkage Between the Broad and Narrow Line Emitting Gas in AGN}

\slugcomment{ submitted May 15, 2002;  accepted July 11, 2002 }

\shorttitle{BLR-NLR Linkage for AGN} \shortauthors{Zamanov et al. }

\author{R. Zamanov\altaffilmark{1}, P. Marziani\altaffilmark{1},
    J. W. Sulentic\altaffilmark{2}, M. Calvani\altaffilmark{1},
    D. Dultzin-Hacyan\altaffilmark{3}, R. Bachev\altaffilmark{2}  }

\altaffiltext{1}{INAF, Osservatorio Astronomico di Padova, Vicolo dell'Osservatorio 5,
   I-35122 Padova, Italy}
\altaffiltext{2}{Department of Physics and Astronomy, University of
   Alabama, Tuscaloosa, AL 35487, USA}
\altaffiltext{3}{Instituto de Astronom\'\i a, UNAM,
   Apdo.Postal 70-264, 04510 Mexico D.F., Mexico}
\altaffiltext{4}{Based in part on data collected at ESO La Silla.}

\begin{abstract}

We investigate the radial velocity difference between the \oiiiopt\ and
\hb\ lines for a sample of $\approx$ 200 low redshift AGN. We identify
seven objects showing an \oiii\ blueshift relative to H$\beta$\ with amplitude
larger than 250 \kms\ (blue ``outliers''). These line shifts are found in
sources where the broad high ionization lines (e.g. \civ)
also show a large systematic
blueshift. Such blueshifts occur only in the population A region of the
Eigenvector 1 parameter domain (that also contains NLSy1 sources).
We suggest that \oiiiopt\ blueshifts are also associated with the
high ionization outflow originating in these sources. This is a direct
kinematic linkage between narrow and broad line emitting gas.

\end{abstract}

\keywords{quasars: emission lines -- quasars: general --  galaxies:
active}

\section{Introduction}

Forbidden \oiiiopt\ emission  arises in
the narrow line region (NLR) of Active Galactic Nuclei (AGN).
This emission has now been partly resolved in the nearest AGN, where
the geometry of the line emitting gas has been
found to be far from spherically symmetric. This suggests that measures of
integrated \oiiiopt\ emission may correlate with source orientation
to the line of sight (Hes et al. 1993; Sulentic, Marziani \& Dultzin-Hacyan 2000a,
and references therein). Observations and theoretical models both (e.g., Steffen
et al. 1997; Sulentic \& Marziani 1999; Moiseev et al. 2001)  suggest a complex
interplay between (where applicable) shocks driven by radio ejection and
the ionizing continuum from the nucleus.

At the same time it is generally believed that radial velocity measures of
the narrow emission lines (e.g. narrow H$\beta$ and \oiiiopt) provide a
reliable measure of the systemic, or rest-frame, velocity. \oiii\ is
preferred because it is not superimposed on a much stronger broad line
component. Limited HI, CO and absorption line measures of the host galaxy
rest frame suggest that \oiiiopt\ usually gives consistent results within
200 \kms\ (de Robertis 1985; Whittle 1985; Wilson \& Heckman 1985; Condon
et al. 1985; Stirpe 1990; Alloin et al. 1992; Evans et al. 2001). Several
observations, however, indicate that the Narrow Line Seyfert 1 (NLSy1)
prototype I Zw 1 shows an \oiiiopt\ blueshift of $\Delta v_r = -$500 \kms\
relative to other rest frame measures (Boroson \& Oke 1987).  This
corresponds to a ~10 \AA\ shift which is larger than any conceivable
measurement or calibration errors; see Marziani et al. 1996).

We report here a study of the velocity shift of \oiii\ relative to \hb.
Our aim was to identify objects with large radial velocity
disagreement between \oiii\ and
\hb\ and their relationship with the general population of AGN.

\section{Sample and Data Analysis}

We measured  the (narrow line) velocity difference between  the peak
of \oiii\ and  of \hb\ using our database of spectra for n= 216 AGN.
The dataset
includes CCD spectra obtained over the past $\approx$10 years for studies
of the \hb\ region in Seyfert 1 and low redshift (z$\la$ 0.8) quasars.
Spectra were obtained with the following telescopes and spectrographs: ESO
1.5m (B\&Ch), San Pedro Martir 2.2m (B\&Ch), Calar Alto 2.2m (B\&Ch), KPNO
2.2m (Gold), and Asiago 1.82m (B\&Ch). Results based on
these spectra can be found in Marziani et al. (1996)
and Sulentic et al. (2000a, 2002).
The unpublished part of this dataset will appear in a forthcoming paper
(Marziani et al. 2002). Spectra were taken with very similar instrumental
setups yielding resolution in the range 4-7 \AA\ FWHM. The S/N of our
spectra is typically in the range $\approx$ 20-40 (minimum 12). The high S/N
and the moderate resolution make this sample
appropriate for the study of line shifts because FWHM and shift measures are not
significantly affected  by undersampling.  Our AGN sample has an average source
absolute B magnitude $<{M_B}> \approx  - 23.7\pm 2.0$ ($H_0 = 50$ \kms
Mpc$^{-1}$, $q_0=0$).

We de-redshifted the spectra using an initial \hb\ measurement for \vr\
in all sources where an \hb\ narrow component could be identified.
Optical \feii\ emission blends were then subtracted using the template
method (Boroson \& Green 1992). At this point, we measured \vr\  for \hb\
a second time, as well as \vr\ for \oiii. Hereafter we will consider only
the difference in radial velocities \dvr\  $= v_r$(\oiii)$- v_r$(\hb) measured
from the \feii\ subtracted spectra. The measurement of \dvr\ turned out to be
possible for 187 sources. The 23 excluded sources include 7 with no
detectable \oiiiopt\ emission  and  16 with a very poorly defined  \hb line
peak.

\section{Results}

\subsection{\oiiiopt\ Line Shift Distribution}

The distribution of \dvr\ measures is shown in Fig. 1a. The values
range from  -950 to +280~km~s$^{-1}$ with an average $<{\Delta v_r}> = -$30
\kms . The sample standard deviation is $\pm$135~\kms .  The median value
is $\overline{\Delta v_r} = - $ 7~\kms . 50\% of the measurements fall in
the interval from $-$49 to $+$16 \kms , so we adopt the first and
third quartile as confidence limits, $\overline{\Delta v_r} =
- 7^{+23}_{-42}$ \kms. The central bins of our \dvr\ distribution are
dominated by measurement errors, with typical values estimated to be in the
range 40--50 \kms\ at 1 $\sigma$ confidence level. The continuous
distribution of measures in the range -200 \kms\ $\la $ \dvr\
$\la$ +200 \kms\ suggests that the \oiii\ redshift measurements are
consistent with \hb\ to within the above range in more then 90\%\ of the
sources. The distribution of shifts is not symmetric around \vr = 0 \kms\ but
is skewed toward the blue. Shifts  in the range -200 $\la $ \dvr $\la$ -100
are three times more frequent than redshifts in the range 100 $\la $
\dvr $\la$ 200 \kms . The maximum shift to the red \dvr $\approx$ 280 \kms\
while shifts up to \dvr $\approx -$1000 \kms\ are observed on the blue
side. The shift distribution reveals seven sources with \oiiiopt\ shift
(\dvr\ $\la$ -250 \kms), which are listed in Table~1.


The relatively rare outliers, (referred to hereafter as `` blue outliers''),
are not randomly distributed in an Eigenvector 1 (=E1) (Sulentic et al. 2000;
\S \ref{e1}) space representation of AGN diversity.  Figure 1b plots \oiii\ shift
amplitude \dvr\ versus full-width half maximum of the \hb\
broad component [FWHM(\hbbc)], and Fig. 1c identifies the blue outlier sources
in the E1 optical plane. Fig. 1b and Fig. 1c suggest that large \oiii\
blueshifts ($\Delta v_r \la$ -300 \kms) are confined to sources with
FWHM(\hbbc) $\la$ 4000 \kms. The largest negative values (\dvr $\la$ -600
\kms) is found at FWHM(\hbbc) $\la$ 2000 \kms. One radio-loud
outlier PKS 0736 ($-$433 \kms\ blueshift) is found. It shows the largest
FWHM(\hbbc) for any of the seven identified extreme blue outliers. Figure
2 presents FeII subtracted spectra of the region of \hb\ and \oiiiopt\
region for the 7  blue outlier sources. They also show preferentially blue
Balmer line asymmetries as was previously noted for NLSy1 sources.

\subsection{Are the Outliers Real ?}


Even though the \oiiiopt\ emission in these sources is weak, it is
difficult to doubt the  detection of such large displacements in {\em both}
[\ion{O}{3}]\l 4959 and [\ion{O}{3}]\l 5007 relative to the  reference
frame defined by \hb\ (Figure 2). We are able to measure the radial
velocity of both \oiii\ and [\ion{O}{3}]$\lambda$4959 including those with
low equivalent width. The velocities are always consistent within 100 \kms
and the equivalent width ratio of \oiii\ and [\ion{O}{3}]$\lambda$4959 is
about 3:1 for  the blue outliers, as expected (see Table 1). This  and the
fact that the shifts as well as the equivalent widths are different from
sources to source, indicates that the \oiiiopt\ lines, although weak, are
not strongly affected by the \feii\ subtraction. We further cheched  with
subtraction of a theoretical template in the region 4800 - 5100 \AA\ (Sigut
\& Pradhan, 2002), and the results, shown in Fig. 2, are very similar.

One of the seven blue outliers (I Zw1) has reliable independent host galaxy
redshift determinations based on HI 21~cm and molecular CO  observations
(see e.g., Sch\"oniger \& Sofue 1994). This is likely to be an accurate and
reliable rest frame determination for at least three reasons: 1) the HI and
CO lines profiles match very closely in shape, width and centroid velocity,
2) the profiles are symmetric and 3) the profiles are steep sided double
horns. The centroid measures are consistent with the velocity of the peak
of the H$\beta$ emission line profile and not with [OIII].

Sources with FWHM(\hbbc ) $\la$ 4000 \kms\ show a sharply peaked,
Lorentzian \hb\ profile. In such cases the narrow line component
is uncertain because no profile inflection is seen. Even if the \hb\ profile
is completely ascribed to the broad component, it seems that the \hb\ peak is
nonetheless a good estimator of the quasar systemic velocity (this result
is expected if the line comes from an extended accretion disk).
We might expect to detect a {\em blueshifted} \hbnc\ analog to the
[OIII] lines, but given the low equivalent width W(\oiii) we normally expect
W(\oiii)/W(\hbnc ) $\sim$ 10), such a blueshifted component
will be lost  in the noise.

We also attempted to identify red outliers (there are  seven
objects with \dvr$\ga$ 150 \kms which form an extended red tail in the shift
distribution). In all cases, we find that the top of
\hb\ is complex or even  multi-peaked. This means that the use of \hb\ as
a reference is sometimes ambiguous. Independent rest frame measures for two of
these sources are consistent with an [OIII] derived radial velocity.
We therefore interpret Figure 1a as consistent with three populations: 1) A scatter
population associated in part with measurement errors;  2) a real, but
relatively rare population of sources where \oiii\ shows an intrinsic large
blue shift relative to the local rest frame (the
blue outliers); 3) an intermediate population, in the range -250 \kms $\la $
\dvr$\la$ -100 \kms), probably containing both sources with intrinsic blue
shifts and sources with larger \hb\ \vr\ uncertainty.

The preferred location of the blue outliers in E1 motivated us to search for
corroboration among samples of phenomenologically similar (see next section) sources.
Grupe, Thomas \& Leighly (2001) identify a blueshift of 570 \kms\ in
RXJ2217-59. Our quick look of other (FeII corrected) spectra (Grupe et al. 1999)
revealed at least three other sources with obvious blueshifts in the range
-300 to -500 \kms\  (RXJ1036-35, RXJ 2340-53 and MS2340-15).

\subsection{Blue outliers and Eigenvector 1 \label{e1}}

Further insight and confidence about the reality of blue outliers can be
gained from considering the distribution in the
Eigenvector 1 diagrams (Sulentic et al. 2000b). The principal physical driver
of E1 is assumed to be the luminosity-to-mass ratio ($\propto$ accretion rate)
convolved with source orientation (Marziani et al. 2001) and/or the black hole mass
(Boroson 2002; Zamanov \& Marziani 2002). The distribution of blue outliers in
the E1 optical plane Fig. 1c) is obviously different from that of the general
AGN population. The blue outliers occupy the lower right part of
the diagram and are exclusively population A/NLSy1 (FWHM(\hbbc)$\la$ 4000 \kms)
sources. A 2D Kolmogorov-Smirnov (see e.g. Fasano \& Franceschini, 1987)
suggests that the  parameter space occupation of the {\em  blue} outliers  is
significantly different than the majority of AGN at a confidence level
of 0.990$-$0.999. The preferred location of the blue outliers in E1 motivated us
to search for other [OIII] blueshifts among samples of phenomenologically similar
sources. Grupe, Thomas \& Leighly (2001) identify a blueshift of 570 \kms\ in
RXJ2217-59. Our quick look of other (FeII corrected) spectra (Grupe et al. 1999)
revealed at least three other sources with obvious blueshifts in the range
-300 to -500 \kms\  (RXJ1036-35, RXJ 2340-53 and MS2340-15).

Figure 3 in Sulentic et al (2000a) involving \civ\ centroid
shift  vs. FWHM H$\beta$ shows a striking similarity to Figure 1b of
this paper. There is a HIL blueshift in the same AGN
(population A) where we find all of the blue outliers. The simplest
interpretation is that we are seeing a kinematic linkage between BLR and
NLR HIL. A source by source comparison reveals 5 of 7 blue outliers with
HST FOS archival spectra of the \civ\ line. The blue outliers show CIV
blueshifts (at FWHM(\civ)) between V$=-$503 and $-$1936 \kms.
This motivates us to look at all sources in our full sample with CIV shift
($>$600 km/s) and W[OIII] ($<$15$\rm{\AA}$), properties similar to the blue
outliers. We find 12 additional sources where: 1) no narrow H$\beta$ peak
and/or [OIII] line is observed (sometimes simply due to noisier than
average spectra), 2) smaller, likely significant,[OIII] blueshifts are
observed (e.g. -200 km/s for PG1444+407) and 3) a peak+blue  wing structure is
observed. In the numerous latter cases we see a narrow unshifted [OIII]
line with a strong blue wing. In the absence of the peak we would measure a
blueshift in the blue outlier velocity range. This motivates us to propose
that the blue outliers represent sources where the classical extended NLR is
absent or suppressed. The peak+ blue wing cases represent sources where both
the (usually strong) classical and the (usually weak) blue shifted NLR
components are present. The concept  of two distinct NLR components is
reenforced by sources like NGC7213 (Busko \& Steiner 1988) and RXJ0148-27
(Grupe et al. 1999) where the unshifted and the blueshifted [OIII]
components are actually resolved.

\section{The Extreme of an Extreme: Large Shifts Governed by Orientation?}

An obvious question involves whether or not the  \oiii\ blue outliers are
peculiar AGN -- a new, previously unknown AGN class. They do not appear to
be part of a continuous distribution of \oiii\ velocity shifts.  However,
even if the distribution of \oiii\ outliers is not consistent with the other
AGN in the E1 parameter plane, blue outliers lie near an extremum in the
AGN occupied domain rather than  outside of it. If we consider our
tentative grid of expected L/M and i values in the E1 plane (Marziani et
al. 2001), we see that the blue outliers may be the product of special
circumstances:  1) a large L/M ratio, and 2) a small inclination
(i$\rightarrow$0$^\circ$). Of course these two circumstances do not explain
the absence of the classical NLR. One possibility is 3) that these sources
are young quasars (see below). It has been suggested that radio-quiet NLSy1
are the equivalent of BL Lacs. We infer that our blue outliers (the ones
observed almost face-on, with largest  \civ\ and \oiii\ blueshifts) may be
the radio-quiet analogous of BL Lac. We note that the single RL blue
outlier identified PKS0736+01 involves an optically violently variable
(OVV) quasar.

We constructed a purely kinematical model in which \oiiiopt\ and \civ\ are
both assumed to arise in a radial flow constrained in a cone of half-opening
angle $\Theta_0\approx 85^\circ$\ (e.g. a high ionization, optically thin
wind) where the receding part of the flow is obscured by an optically thick
accretion disk (i.e., we are able to see the approaching part of the flow
and the near side of the disk which is assumed to emit \hb).  We assumed
that radial motions in the gas  were a fixed fraction (1.5) of the local
virial velocity.  We integrated over a region   100 $R_g \le r \le 10^5
R_g$\  for \civ\ and over a region 300 $R_g \le r \le 10^5 R_g$ \ for
\oiii, with emissivity power law-indices q = -2 (\civ) and q = -1 (\oiii).
Fig. \ref{MODEL} shows the resulting model profiles for a viewing angle
i$\approx$ 15$^\circ$. The width and shift  of both lines are  accounted
for. We suggest, without considering this particular model as a physical
one, that emission from outflowing gas, possibly associated to a disk wind,
can explain the observed profiles. The occurrence of large \oiii\  shifts
seems to be associated with low W(\oiii). Our model integration also
consistently suggests {\em a very compact} NLR (r$\sim$ 1 pc for a black
hole mass 10$^8$\ solar masses).  In this case, the receding part of the
outflow may be more easily hidden by an optically thick disk extending to
$\sim$ 1 pc (this is a requirement also to avoid double peaked profiles).
The filling factor needed to explain the \oiii\ luminosity of I Zw 1 and PG
1543+489 can be reasonably small, $\sim 10^{-3}$. It is intriguing that a
compact NLR complements several lines of evidence suggesting that NLSy1s
are young AGN. Large W(\oiii) would imply a larger emitting volume and
hence may be associated with lower \oiii\ shifts. The frequent observations
of blueward asymmetry close to the \oiiiopt\ line profile base indicates
that the same outflow may also be occurring in the innermost NLR of AGN
with larger W(\oiii).

It is also interesting to note that Broad Absorption Line QSOs (BAL QSOs)
are found more frequently in samples with low W(\oiiiopt) (Boroson \&
Meyers 1992; Turnshek et al. 1997). We do not find any BAL QSO among the
outliers. In addition, the known low-z BAL QSO in our sample (PG 1004+13, PG
2112+059) with measurable \oiii\ (2 out of 5) show shifts \dvr\ $\sim$ 0
\kms. This is also  consistent with orientation playing a role and with
blue outliers being oriented predominantly ``face-on.''  On the contrary,
the \civ\ absorption/emission profiles of BAL QSOs suggest that they may be
observed far from pole-on (Marziani et al. 2002, in preparation).

\section{Conclusion}

We find that the  \hb\ and \oiii\  lines provide measures of the radial
velocity of AGN usually consistent within $\pm$200 \kms. We identified
infrequent ($\approx$ 5\% in a sample of 200 sources)  AGN showing \dvr
$\la$ -250 \kms. They belong to the  extreme population A sources that also
show a large broad line \civ\ blueshift. This kinematic coupling of the NLR
and BLR HIL emission most likely involves a wind or outflow. Our analysis
suggests that blue outliers are not peculiar objects, but rather AGN viewed
of extreme L/M with a compact NLR. 
A predominance
of blueshifts in the sample indicates that \oiii\ peak velocity is affected
by outflow motions occurring in the innermost NLR.

\begin{acknowledgements}
We are very grateful to Giovanna Stirpe  for fruitful discussions. The
authors acknowledge  support from the Italian Ministry of University and
Scientific and Technological Research (MURST) through grant and Cofin
00$-$02$-$004.
This research has made use of the NASA/IPAC
Extragalactic Database (NED) which is operated by the JPL, Caltech under
contract with the NASA.
\end{acknowledgements}

\begin{figure*}
\vspace{8.0cm}
\includegraphics{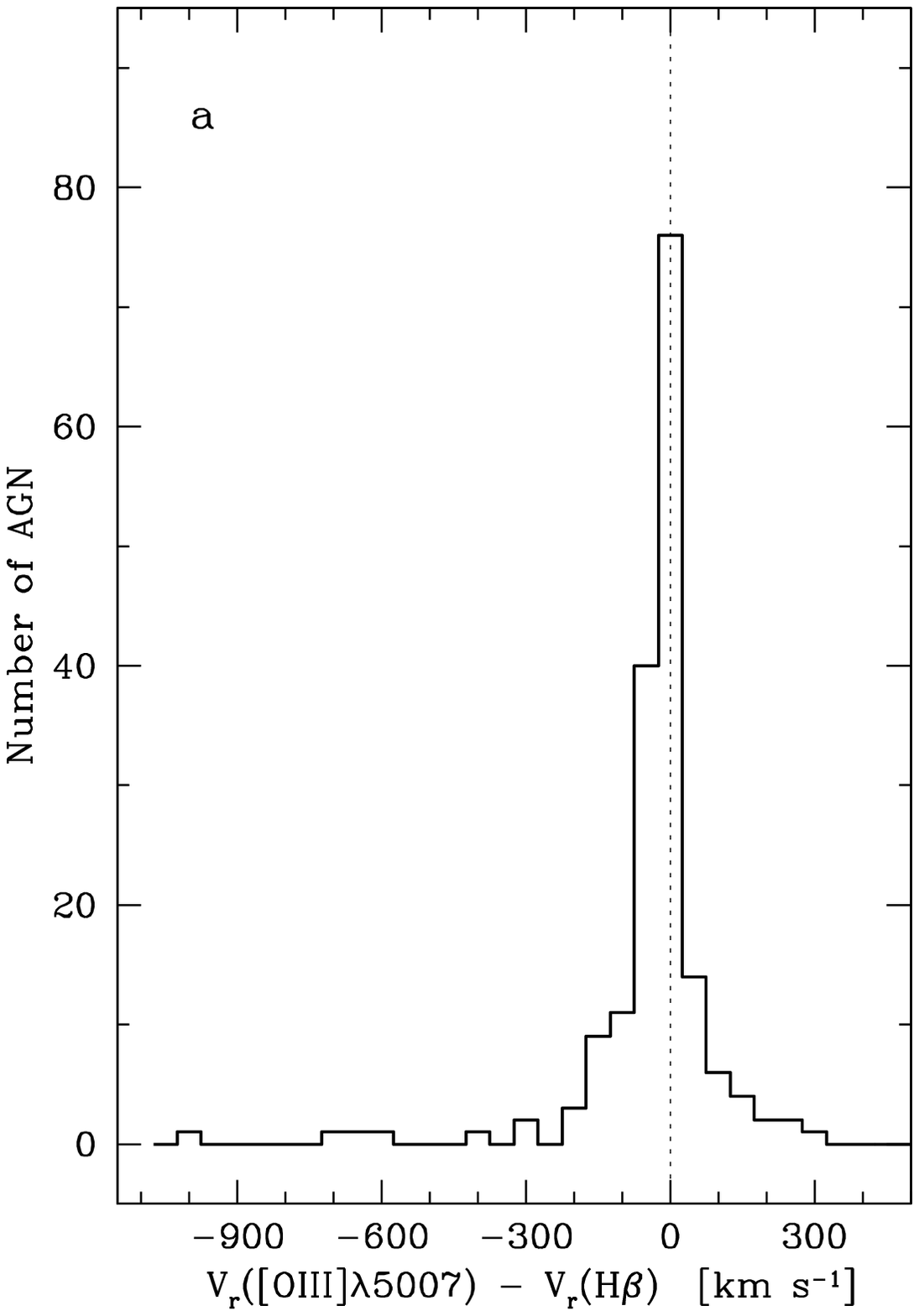}
\includegraphics{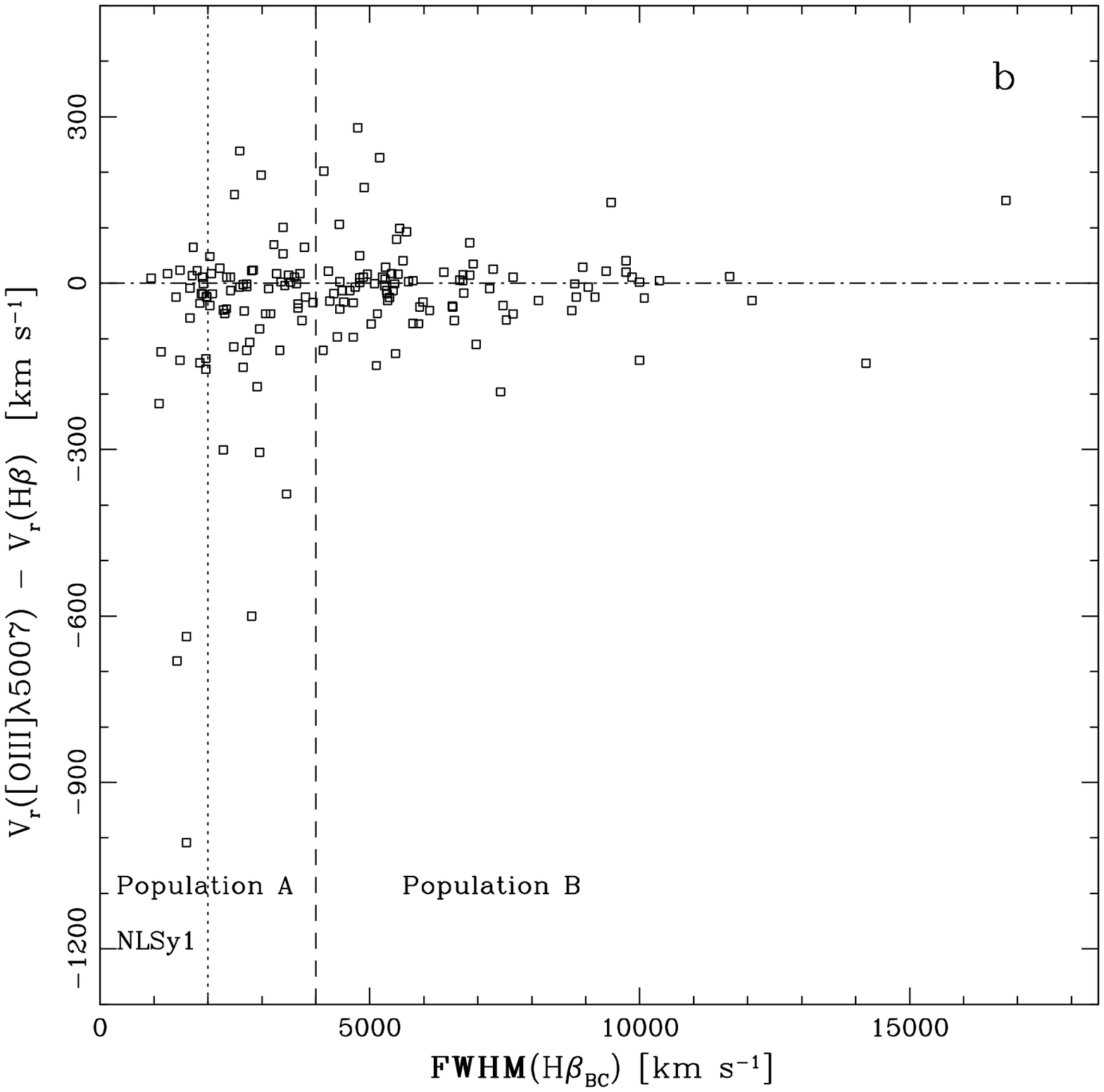}
\includegraphics{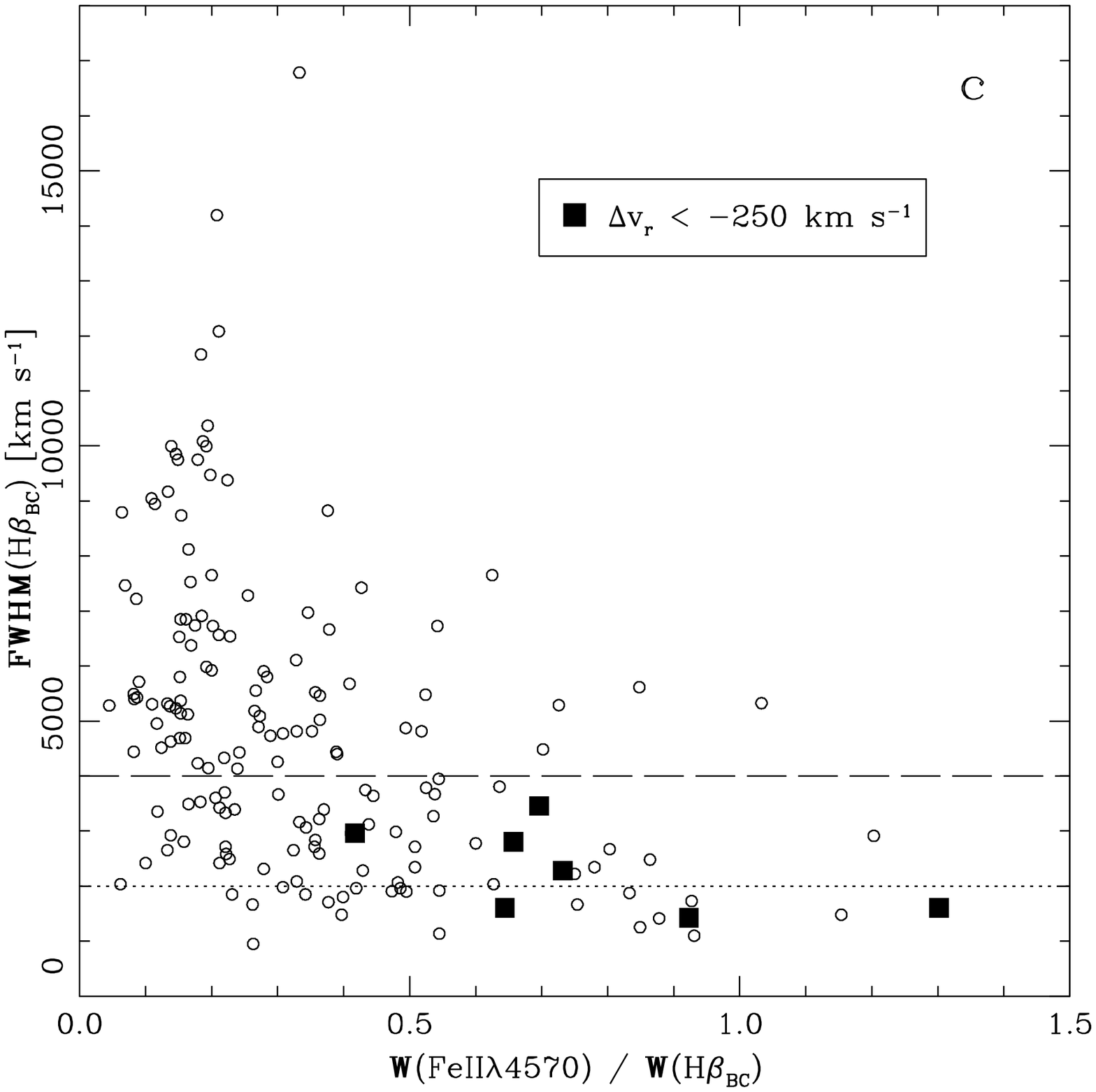}

\caption[]{
 {\bf (a)} Histogram showing the distribution of the radial velocity difference
 between the \oiii\ and the \hb\ line.
 {\bf (b)} The difference of the radial velocity between the \oiii\ and \hb\
 versus FWHM(\hbbc).
 The vertical dotted line marks the boundary of the NLSy1
 galaxies. The vertical dashed line separates Population A and B sources.
  Seven objects with
 with \dvr$ < -250$~\kms\  are visible.
 {\bf (c)} Location of outliers in the FWHM(\hbbc)
 vs. \rfe\ diagram (the optical Eigenvector-1 diagram).
 Filled squares represent the ``blue outliers'' (\dvr
 $<-250$~km~s$^{-1}$).}
\label{FWHM}
\end{figure*}

\begin{figure}
\includegraphics[height=14.0cm, angle=0]{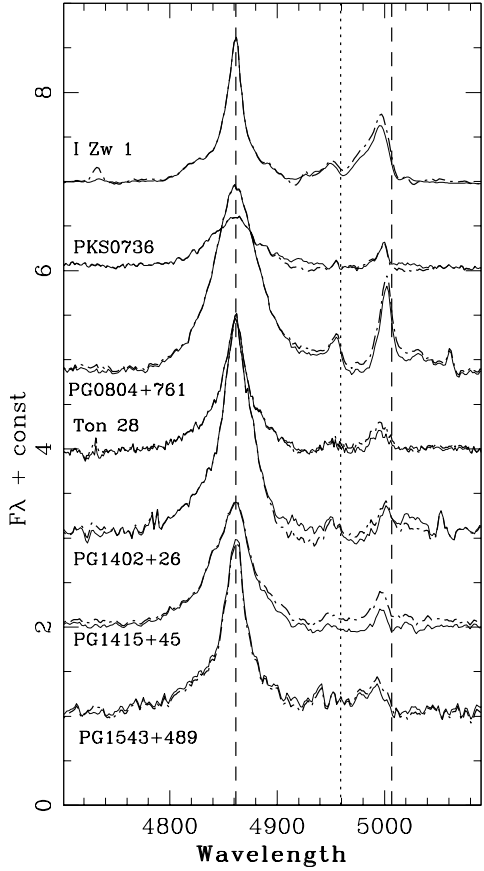}
\caption{ \hb\ spectral region of the ``blue'' outliers   after the
deredshift and   subtraction of the FeII template.
   The spectra are normalized with respect to the local continuum and arbitrary constant added. The solid line corresponds to the subtraction of
   a I Zw 1-based empirical template, the dot-dashed line  to the subtraction of a theoretical template.
   The vertical lines indicate the position of \hb, [OIII]$\lambda$4959 and [OIII]$\lambda$5007.
   The difference  in the radial velocities between the \oiii\ lines and  \hb\ is well visible.} \label{SPEC}
\end{figure}

\begin{figure}
\includegraphics[width=8.0cm, height=8.0cm, angle=0]{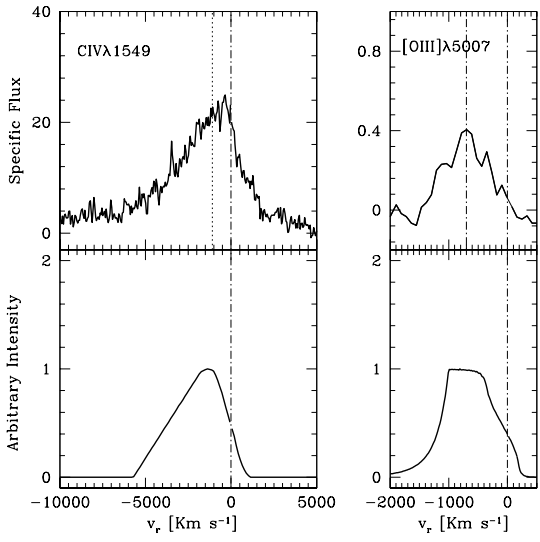}
\caption{ Upper panel: the \civ\ and \oiii\ profile of Ton 28,
    a blue outlier. Lower Panel: \civ\ and \oiii\ outflow model
    profiles, for optically thin gas moving at approximately the local
    escape velocity. Profiles have been computed for a cone of
    half-opening angle 85$^\circ$, with the line of sight oriented at
    15$^\circ$~ with respect to the cone axis. The receding part of the
    flow is assumed to be fully obscured from the observer.}
\label{MODEL}
\end{figure}

\begin{centering}
\begin{table*}[]
\caption{ Objects with anomalous difference of the radial velocities
      between the [OIII]$\lambda$5007 and H$_{\beta}$ (``blue outliers"). }

\begin{tabular}{cccccccc}
\hline

 Name       &  z(H$\beta$)   &   \dvr\    & W[OIII]       &  W[OIII]      & \rfe &  FWHM(\hbbc) & $\Delta$ \vr(\civbc) \\
            &                &            & $\lambda$4959 & $\lambda$5007 &      &          &       \\
        &                &  [\kms]    & [ \AA\ ]      &  [ \AA\ ]     &      &  [\kms]  & [\kms]        \\
\hline

I Zw 1      & 0.0606 &-640$\pm$30 & 4.3 &  15.3 & 1.30 &  1600  & -820     \\
PKS 0736+01 & 0.1909 &-430$\pm$60 & 0.6 &   2.6 & 0.70 &  3460  & \dots \tablenotemark{a} \\
PG 0804+761 & 0.1014 &-305$\pm$30 & 3.3 &  10.1 & 0.42 &  2960  & \dots \tablenotemark{a} \\
Ton 28      & 0.3297 &-680$\pm$50 & 1.6 &   3.4 & 0.71 &  1860  & -1120  \\
PG 1402+261 & 0.1651 &-300$\pm$50 & 1.4 &   2.6 & 0.73 &  2280  & -650     \\
PG 1415+452 & 0.1151 &-600$\pm$50 & 1.1 &   2.9 & 0.66 &  2810  & -950    \\
PG 1543+489 & 0.4009 &-950$\pm$50 & 2.3 &   6.5 & 0.64 &  1600  & -2630  \\

\hline
\end{tabular}
\tablenotetext{a}{ HST/FOS observations not available.}
\tablecomments{ The typical errors are
$\pm$30\% in W$[\lambda 4959]$,
$\pm$10\% in W$[\lambda 5007]$,
$\pm$0.15 in \rfe,
$\pm$150 \kms\ in FWHM(\hb),
and $\pm$200 \kms\ in  $\Delta$ \vr(\civbc).
}

\end{table*}
\end{centering}

\begin{thebibliography}{}

\bibitem[Alloin1992]{1992A&A...265..429A}  Alloin, D., Barvainis, R.,
             Gordon, M.~A.,  \& Antonucci, R.~R.~J.\ 1992, \aap , 265, 429.
\bibitem[2002]{B2002}     Boroson T., 2002, ApJ 565, 78
\bibitem[2001]{BG1992}    Boroson T.A., Green R.F., 1992, ApJS, 80, 109
\bibitem[2001]{BM1992}    Boroson T.A.,\& Meyers, K.~A., 1992, ApJ, 397, 442
\bibitem[1987]{BO1987}    Boroson, T.~A.~\& Oke, J.~B.\ 1987, \pasp, 99, 809.
\bibitem[1988]{BS1988}    Busko, I. ~\& Steiner, J. 1988,\mnras, 232, 525.
\bibitem[Condon1985]{1985AJ.....90.1642C} Condon, J.J., Hutchings, J.B., \& Gower, A.C.\ 1985, \aj, 90, 1642.
\bibitem[de Robertis(1985)]{1985ApJ...289...67D} de Robertis, M.\ 1985, \apj, 289, 67.
\bibitem[2001]{Ev2001}    Evans, A.~S., Frayer, D.~T., Surace, J.~A., \& Sanders, D.~B.\ 2001, \aj, 121, 1893.
\bibitem[Hes, Barthel, \& Fosbury(1993)]{H1993} Hes, R., Barthel, P.~D., \& Fosbury, R.~A.~E.\ 1993, \nat, 362, 326
\bibitem[2001]{Fasano}    Fasano G., \& Franceschini A., 1987, MNRAS 225, 155
\bibitem[1999]{Grup1}     Grupe, D., Beuermann, K., Mannheim, K., Thomas, H.-C., 1999, \aap, 350, 805
\bibitem[2001]{Grup2}     Grupe, D., Thomas, H.-C., Leighly, K. M., 2001 \aap, 369, 450
\bibitem[1996]{M1996}     Marziani, P., Sulentic, J. W., Dultzin-Hacyan, D., Calvani, M., Moles, M., 1996, ApJS, 104, 37
\bibitem[2001]{M2001}     Marziani, P., Sulentic, J.W., Zwitter, T., Dultzin-Hacyan, D., Calvani, M., 2001, ApJ, 558, 553
\bibitem[Mirab]{Mirab}    Mirabel I.F., 1992, in Relationships Between Active Galactic
   Nuclei and Starburst Galaxies, ed. A. Filippenko, ASP Conf.Ser. vol. 31, p. 347
\bibitem[2001]{Mo2001}    Moiseev, A.~V., Afanasiev, V.~L., Dodonov, S.~N., Mustsevoi, V.~V., \& Khrapov, S.~S.\
   2001, IAU Colloq.~184: AGN Surveys, E79
\bibitem[2001]{Rich2001}  Richter, P., Savage, B.D., Wakker, B.~P., Sembach, K.~R., Kalberla, P. M. W., 2001, ApJ, 549, 281
\bibitem[1994]{ss94}      Sch\"oniger, F., \& Sofue, Y. 1994, \aa, 283, 21
\bibitem[2002]{Sig2002}   Sigut T.A.A., Pradhan A.K., 2002,  ApJS, in press (astro-ph/0206096)
\bibitem[1997]{St1997}    Steffen, W., Gomez, J.~L., Raga, A.~C., \& Williams, R.~J.~R.\ 1997, \apjl, 491, L73
\bibitem[1990]{G.Stirpe}  Stirpe, G.~M.\ 1990, \aaps, 85, 1049.
\bibitem[1999]{SM99}      Sulentic, J.~W.~\& Marziani, P.\ 1999, \apjl, 518, L9
\bibitem[2000]{Su2000a}   Sulentic J.W., Marziani P., \& Dultzin-Hacyan D., 2000a, AR A\&A, 38, 521
\bibitem[2000]{Su2000b}   Sulentic, J.~W., Zwitter, T., Marziani, P., \& Dultzin-Hacyan, D.\ 2000b, \apjl, 536, L5.
\bibitem[2002]{su2002}    Sulentic, J.~W., Marziani, P., Zamanov, R., Bachev, R., Calvani, M. \& Dultzin-Hacyan, D. \
2002, \apjl, 566, L71
\bibitem[1985]{Whittle}   Whittle, M.\ 1985, \mnras, 213, 33.
\bibitem[1985]{WH85}      Wilson, A.~S.~\& Heckman, T.~M.\ 1985, Astrophysics of Active Galaxies and Quasi-Stellar
     Objects, Mill Valley, CA, University Science Books, 1985,  39
\bibitem[2002]{ZM2002} Zamanov, R., \& Marziani, P., 2002, ApJ, 571, L77

\end{thebibliography}
\end{document}